\documentstyle[aas2pp4]{article}
     
\slugcomment{Submitted to {\em The Astrophysical Journal}}

\lefthead{PAGE \& SARMIENTO}
\righthead{MAGNETIZED NEUTRON STAR TEMPERATURE. II.}

\newcommand{\be}{\begin{equation}} 
\newcommand{\ee}{\end{equation}} 

\newcommand{\ba} {$\begin{rm}\begin{array}{c} }
\newcommand{\bba} {$\left\{ \begin{rm}\begin{array}{c} }
\newcommand{\baa}{$\begin{rm}\begin{array}{cc}}
\newcommand{\ea} {\end{array}\end{rm}$}
\newcommand{\eab} {\end{array}\end{rm} \right.$}

\newcommand{\Msol}{\mbox{$M_{\odot}\;$}}
\newcommand{\Msun}{\mbox{$M_{\odot}\;$}}

\begin{document}
 
\title{SURFACE TEMPERATURE OF A MAGNETIZED NEUTRON STAR AND
       INTERPRETATION OF THE {\em ROSAT} DATA. II.}
\author{Dany Page and A. Sarmiento}
\affil{Instituto de Astronom\'{\i}a, UNAM, Apdo. Postal 70-264,
       04510 M\'{e}xico D. F., M\'{e}xico.}

\begin{abstract}

We complete our study of pulsars' non-uniform surface temperature and of its
effects on their soft X-ray thermal emission.
Our previous work had shown that, due to the effect of gravitational
lensing, dipolar fields cannot reproduce the strong pulsations observed in
the four nearby pulsars for which surface thermal radiation has been
detected: PSR 0833-45 (Vela), PSR 0656+14, PSR 0630+178 (Geminga), and PSR 1055-52.
Assuming a standard neutron star mass of 1.4 \Msol,
we show here that the inclusion of a quadrupolar component, if it is suitably
oriented, is sufficient to increase substantially the pulsed fraction, $Pf$,
up to, or above, the observed values if the stellar radius is 13~km or even 10~km.
For models with a radius of 7~km the maximum pulsed fraction 
obtainable, with (isotropic) blackbody emission, is of the order of 15\%
for orthogonal rotators (Vela, Geminga and PSR 1055-52) and only 5\% for an inclined
rotator as PSR 0656+14. 
Given the observed values, this may indicate that the neutron stars in Geminga
and PSR 0656+14 have radii significantly larger than 7~km and, given that
very specific quadrupole components are required to increase $Pf$, even radii of the
order of 10~km may be unlikely in all four cases.
However, effects not included in our study may possibly seriously invalidate this 
temptative conclusion.

We confirm our previous finding that the pulsed fraction always
increases with photon energy, below about 1 keV, when blackbody
emission is used and we show that it is due
to the hardenning of the blackbody spectrum with increasing temperature.
The observed decrease of pulsed fraction may thus suggest that the emitted spectrum
softens with increasing temperature and that this observed effect must be of
atmospheric origin.

Finally, we apply our model to reassess the magnetic field effect on
the outer boundary condition used in neutron star cooling models and
show that, in contradistinction to several previous claims, it is very small 
and most probably results in a slight reduction of the heat flow
through the envelope.

\vspace{1.cm}
\noindent
Submitted to {\em The Astrophysical Journal}.

\end{abstract}


\section{INTRODUCTION 
         \label{sec:intro}}

In a previous paper (Page 1995a; hereafter Paper I) we had presented a simple
but realistic model of temperature distribution at the surface of a 
magnetized neutron star.
This model was used to study the effects of temperature inhomogeneity
on the neutron star thermal emission and compare them with the
observed properties of the four neutron stars for which this emission
has been detected by the {\em ROSAT} satelite (\"Ogelman 1995a; Page 1995b):
PSR 0833-45 (Vela; \"Ogelman, Finley, \& Zimmermann 1993),
PSR 0656+14 (Finley, \"Ogelman, \& Kizilo\mbox{\u{g}}lu 1992),
PSR 0630+178 (Geminga; Halpern \& Holt 1992) and
PSR 1055-52 (\"Ogelman \& Finley 1993).
As a first step we did not take into account magnetic effects in the 
atmosphere which are also very large.
We also restricted ourselves to dipolar magnetic fields and the three main
results we obtained were:
\newline $-$ Magnetic effects on heat transport in the neutron star envelope
do induce very large temperature differences at the surface but when
gravitational lensing is taken into account and only dipolar fields are
considered the predicted pulsed fractions are smaller than those observed in
the soft band where the surface thermal emission is seen. 
In particular, in the case of 1.4 \Msol neutron stars with
small radii $\sim 7.0 - 9.0$ km, the pulsed fractions $Pf$s are below 1\%.
With very small radii $ \leq 6.5$ km, and the same mass, the
gravitational beaming can however increase $Pf$ up to 6 - 7\%.
Observed values of $Pf$ range between 10 - 30\%.
\newline $ - $ With dipolar fields the observable light curves are
very symmetrical and their shapes do not correspond to the 
observations. 
Together with the previous result, this lead us to conclude that the inclusion 
of dipolar fields only is not sufficiently adequate to model the surface 
magnetic field of the four observed neutron stars.
\newline $ - $ The amplitude of the pulse profile increases with
increasing photon energy.
This result does not correspond to what is observed in the soft X-ray band
in the cases of PSR 0656+14, Geminga and PSR 1055-52.

We complete here this study and consider also the effects of these 
surface temperature distributions on the cooling of neutron stars.
Since the completion of the work presented in Paper I, complementary
results considering the magnetic effects in the atmosphere, where the
emerging spectrum is generated, have been presented (Pavlov {\em et al.} 1994;
Zavlin {\em et al.} 1995)
and clearly showed that they can be as important as the
magnetic effects in the envelope that we consider here and in Paper I.
A complete study must obviously include both aspects of the problem, envelope
and atmosphere.
However, as long as the exact chemical composition of a pulsar's
surface is not known, analysis with blackbody (BB) emission will still
remain a mandatory first step and a reference to which atmosphere
models will be compared.
For this reason it is important to characterize BB emission and
determine what can and what cannot be obtained with it.
As important as the successes of our model will thus be its failures which can
guide us toward the correct atmosphere, or more generally surface, model
(Page 1995b; Page, Shibanov, \& Zavlin 1995).
Nevertheless, several of our results are, we hope, sufficiently general and
robust to be valid despite of the limitations of BB emission.
We restrict ourselves to study general properties of the model and refrain
from detailed analyses of the data which we leave for future work.
Our method to generate surface temperature distributions and the observable
X-ray fluxes is described in detail in Paper I.

In \S~\ref{sec:data} we present some complements to our summary of the {\em ROSAT}
data given in Paper I.
In \S~\ref{sec:quadr} we consider the effects of the inclusion of 
the quadrupolar components of the magnetic field and show that when 
they are superposed to a dipolar field they provide a sufficiently general 
configuration for our study.
Some basic and straightforward results are presented in \S~\ref{sec:strength}
and \ref{sec:spectra}.
We use these configurations to discuss the reliability of our
surface temperature model in \S~\ref{sec:reliability} and also to
study the possibility of putting constraints on the neutron star (NS)
size through gravitational lensing in \S~\ref{sec:lens}.
We then discuss the energy dependence of the light curves amplitude in
\S~\ref{sec:pf}, this section will give a clear indication of the inadequacy
of BB emission for understanding detailed features of the observations.
The next section, \ref{sec:caps}, adds a second component of
emission in order to model the higher energy tail of Geminga.
Section~\ref{sec:cooling} studies the effect of the surface temperature
distribution on the boundary condition used for modeling the thermal evolution
of neutron stars.
Finally, \S~\ref{sec:concl} presents our conclusions.

\section{THE {\em ROSAT} DATA
          \label{sec:data}}

Since the completion of Paper I some information has been presented
on the second set of {\em ROSAT} observations of PSR 0656+14 (\"Ogelman 1995a;
Possenti, Mereghetti, \& Colpi 1996).
The complete set of data clearly shows the presence of
a hard tail which is strongly pulsed, as in Geminga and PSR 1055-52.
Both the energy dependence of the pulsed fraction and of the pulse phase
show some similarity with PSR 1055-52 (\"Ogelman 1995a) and also
with Geminga (\"Ogelman 1995b):
in all three cases $Pf$ is almost constant up to channels $\sim 40$ ($\sim 30$
for Geminga), then decreases and finally increases (very strongly
in the case of PSR 1055-52).
Moreover, at around the same energy where $Pf$ increases, the phase of the
peak changes.
The spectral fits with a soft BB for the surface thermal emission and a BB
or power law for the hard tail show that the surface thermal emission dominates
at channels below channel $\sim 50$ in the case of Geminga 
(e.~g., figure~\ref{fig:Geminga}) and channel $\sim 100$ in the cases of PSR 0656+14
and 1055-52 (see, e.~g., \"Ogelman 1995a).
An important point is that in the case of Geminga the change in the
peak phase coincides with the spectral shift from the surface thermal
emission to the hard tail (see \S~\ref{sec:caps}) and is not surprising.
However, in the two other cases this phase shift apparently, and
intriguingly, occurs within the band dominated by the surface thermal
emission.

\section{ QUADRUPOLAR FIELDS 
          \label{sec:quadr}}

\subsection{The quadrupolar components and gravity effects
            \label{subsec:quadr}}

The first natural step beyond dipolar fields is the inclusion of a
quadrupolar component which we write as
\be
{\bf B}^Q = \sum_{i=0,4} Q_i \left(\frac{R}{r}\right)^4{\bf b_i}
\label{equ:quadr}
\ee
where the five generating fields $\bf b_i$ are listed in 
table~\ref{tab:quadr}.
Considering the generating quadrupole components separately at the
star's surface, one finds that all components reach a maximum strength
of 1 and ${\bf b}_0$ has a minimum of $1/\sqrt 5$ (=0.45) while the other
four have a minimum of zero.
They have four magnetic poles except for $\bf b_0$ whose `south' pole
is degenerate and becomes a line covering the whole equator. 
However, the general quadrupolar field ${\bf B}^Q$ can have up to six
poles, south and north poles always being in an even number. 
When the quadrupole is added to a dipole we can have again up to six
poles, and the number of north (south) poles can be odd.
Notice finally that the scale value of the dipolar component
we use is the field strength at magnetic pole, i.~e., twice the
value commonly considered, so that all field scales we cite always refer
to the maximum field strength of the cited component 
at the stellar surface in flat spacetime.

We also include the effect of gravity on the magnetic field.
It is not as important as red-shift and lensing but is nonetheless not
completely negligible since it can substantially increase the field strength
and is moreover straightforward to implement.
It can be written in the form of four multiplicative
factors, two for the radial component of the magnetic field and two
for its angular components:
if $B_r$, $B_\tau$ ($\tau = \theta$ or $\phi$) denote the radial and angular
components of the magnetic field in the absence of a gravitational field, then: 
\be 
B^D_{r g}    =                  f_1 \; B^D_r,     \qquad \qquad 
B^D_{\tau g} = \sqrt{g_{00}} \; g_1 \; B^D_\tau 
\ee 
give the dipolar fields in the presence of a
gravitational field, and: 
\be 
B^Q_{r g}      =                f_2 \; B^Q_r,     \qquad \qquad
B^Q_{\tau g} = \sqrt{g_{00}} \; g_2 \; B^Q_\tau 
\ee
give the quadrupolar fields in the same situation. The factors in the previous
equalities are given by: 
\be 
f_1 = - \frac{3}{x^3} \left[\ln(1-x) + \frac{1}{2}x(x+2)\right]
\ee
\be
g_1 = -2 f_1 + \frac{3}{1-x}
\ee 
\be 
f_2 = \frac{10}{3 x^4}
      \left[ 6 \ln(1-x) \frac{(3x-4)}{x} + x^2 + 6 x - 24 \right]
\ee
\be
g_2 = \frac{10}{x^4}
      \left[ 6 \ln(1-x) \frac{2-x}{x} + \frac{x^2-12 x + 12}{1-x} \right]
\ee
where $x = R_S/r$, $R_S = \frac{2 G M}{c^2}$ being the star's Schwarzschild radius,
and $g_{00} = 1-x$ is the time component of the metric
(Muslimov \& Tsygan 1987).
These corrections of course tend towards 1 at large distances.
The values of $f_l$ and $g_l$ at the stellar surface, i.~e., at
$x = R_S/R$, are plotted in figure~\ref{fig:corr}: 
the effect is stronger on the quadrupole than on the dipole, simply
due to the stronger $r$ dependence of the former.
If $R \rightarrow R_S$ then $f_l$ and $g_l$ tend to $\infty$ which
means that, for a given surface field, the field `at infinity'
vanishes: this is a particular case of the `no hair' theorem for non
rotating blackholes (see, e.~g., Misner, Thorne, \& Wheeler 1970).

\subsection{Statistics of dipole+quadrupole fields
            \label{subsec:stat}}

Since the quadrupolar component introduces five more degrees of freedom,
a clear general assessment of its possible effects is necessary.
To investigate this, we computed nine sets of 1000 models each, in
which the quadrupolar components where chosen at random by the computer:
three orientations of the dipole and observer, $\alpha = \zeta = 30^\circ$, 
$60^\circ$ and $90^\circ$, were taken to test for geometrical effects 
($\alpha$ is the angle between the rotation axis and the dipole
and $\zeta$ between the rotation axis and the observer's direction)
and three values for the stellar radius, $R$ = 7~km,
10~km and 13~km, were used to analyze the effect of gravitational lensing.
The maximum lensing angles $\theta_{max}$ are respectively,
$194^{\circ}$, $132^{\circ}$, and $117^{\circ}$ for radii of 7, 10, and 13 km,
respectively (figure 3 in Paper I).
The dipole strength was fixed at 10$^{12}$ G (value at the magnetic
pole in flat space-time; GR effects increase it slightly) and the 1000
quadrupoles were added to the basic dipole.
The quadrupole component strengths, $Q_i$, were restricted to the range
$10^{10}$~G -- $5~\cdot~10^{12}$~G.
The star's surface temperature was calculated using an interior temperature
of $T_b$ = 10$^8$ K which implies an effective temperture around
$10^6$~K (the exact value depending on the star size and field configuration).
Notice that GR effects depend only on the ratio $R/R_S = Rc^2/2GM$ of the star's
radius $R$ to its Schwarzschild radius $R_S$, so that all our results can be extrapolated to
other masses and radii (see, e.~g., figure~\ref{fig:corr} for easy conversion).
To save CPU time we only calculated the variation of the observable
phase dependent effective temperature, $T_e^{\Phi}$ (Paper I, \S~5.2),
during the star's rotation, i.~e., the effective temperature of the
portion of the stellar surface visible to the observer at a given
phase $\Phi$.
Selecting a subset of the 9,000 models so generated we performed complete
calculations of the detectable fluxes to calibrate the observable pulsed
fraction $Pf$, in the PSPC's channel range 7~--~50, in terms of the pulsation
of $T_e^\Phi$.
The results of these calculations are shown in figure~\ref{fig:stat} where
each model is shown as a black dot and the pure dipole case
as a black-white ring.
The dramatic effect of lensing is again clearly seen.
For later use, figure~\ref{fig:stat} plots the ratio of the star's effective
temperature $T_e$(mag.) for the given field configuration to the
effective temperature it would have without the magnetic field,
$T_e$(non mag.), at the same interior temperature $T_b~=~10^8$~K.
A different $T_b$ would change the values of $Pf$ but not change
significantly the statistics for the number of models giving large $Pf$.
The observable $Pf$ increases with decreasing temperature for a given
surface magnetic field configuration as explained in \S~\ref{sec:strength}.

We want here to correct our definition of the pulsed fraction (Eq. 21 in
Paper I) and rather use its standard expression as the fraction of counts
above minimum, thus
\be
Pf(i) = \frac{(Cts_{mean} - Cts_{min})}{Cts_{mean}}
\ee
where $Cts_{mean}$ and $Cts_{min}$ are, respectively, the mean and minimum
count rates detected during the star's rotation in channel \#~i.
For sinusoidal pulse profiles this is equivalent to our former definition
and does not affect the results of Paper I.

The first obvious and natural result of figure~\ref{fig:stat} is
that most configurations produce little pulsation: 
most quadrupoles induce several warm regions which
flatten the pulse profile even below the value of the pure 
dipole case.
However, many configurations do rise $Pf$ substantially,
compared to the purely dipolar case,
and to a high value in some special cases.
We can thus obtain observable pulsed fractions higher than 30\%
(at energies below 0.5 keV and at $T_e \sim 10^6$~K)
for a star of radius 13 km.
For stars of radii 10 km a $Pf$ value comparable to the observed
ones can also be obtained.
The most interesting case is the 7 km radius star: despite of the
enormous gravitational lensing in this star ($\theta_{max} = 194^\circ$), 
we can still find a configuration which gives a $Pf$ above 10\% at high 
$T_{e}$ ($\sim 10^6$ K) and above 15\% at $T_{e} \sim 3 \cdot 10^5$ K, as 
we have checked explicitly.

A last point is worth mentioning: the strong increase in the pulsed fraction
possible with the addition of a quadrupolar component is {\em not} due to an
increase in the area of the cold regions, compared to the purely dipolar case,
but rather to a displacement of the magnetic poles and the surrounding warm regions
which brings them closer to each other (see, e.g., figure~\ref{fig:model1-temp}).
The pulse profiles of Geminga and PSR 1055-52, which are both
considered to be almost orthogonal rotators, show a single wide peak
which indicates that the two polar caps and the surrounding warm regions
are much less than 180$^\circ$ apart.
How close the magnetic poles can be pushed toward each other is what
will determine the `success' of the quadrupolar component in
increasing the pulsed fraction, whereas quadrupoles which moreover
introduce several new warm regions will fail.
Figure~\ref{fig:stat} illustrates this clearly: the ratio
\mbox{$T_e$(mag)/$T_e$(non mag.)} is practically the same for the
strongly pulsed dipole+quadrupole configurations as for the purely
dipolar case, about 0.9, while weakly pulsed configurations which
induce several warm regions may have a ratio larger than one.
These considerations will be of importance for \S~\ref{sec:cooling}.

\section{EFFECT OF THE FIELD STRENGTH
            \label{sec:strength}}

Considering the effect of the field strength, in the purely dipolar
case we found that the maximum $Pf$ is obtained when the overall
surface field reaches a few times $10^{11}$~G (Paper~I): the same
result still applies to the dipole + quadrupole case.
Below \mbox{$2 - 3 \cdot 10^{11}$ G} the magnetic effects are getting smaller
and above this value, even if the magnetic effects are increasing,
the resulting observable $Pf$ varies very little.
The reason why the pulsed fraction depends very weakly on the field
strength for large enough fields is simple:
cold regions contribute so little to the total flux that it does not
really matter how cold they are, but only how large they are.
We are dealing mostly with warm plates whose extension is determined
more by the field orientation than by the field strength as long as the field
strength does not drop below a few times 10$^{11}$ G.

\section{EFFECT OF THE TEMPERATURE DISTRIBUTION ON THE SPECTRUM
            \label{sec:spectra}}

The surface temperature distributions induced by dipole + quadrupole fields 
can have quite complicated structures. 
However, as discussed in Paper I for dipolar fields, the resulting phase
integrated spectra are close to single temperature spectra at the corresponding
effective temperature.
Figure~\ref{fig:comp-spectrum} shows examples at five temperatures for
the dipole+quadrupole field from the set generated for figure~\ref{fig:stat}
(see \S~\ref{subsec:stat}) which gives the highest pulsed fraction.
The surface temperature distribution induced by this field configuration
is illustrated in figure~\ref{fig:model1-temp}.
The result is very close to the effect of a simple dipole (figure 6 in
Paper~I) due to the fact that the chosen quadrupole induces warm
regions of approximately the same size as the pure dipole but at different locations
(see \S~\ref{subsec:stat}).
When taking into account the PSPC's response (figure~\ref{fig:comp-spectrum}~B)
the composite spectrum presents a clear excess compared to the single temperature
spectrum at channels above channels 50~--~60.
This excess will not affect spectral fits in the case of Geminga since the flux
at channels above 60 is dominated by the hard tail.
However, in the three other cases where the hard tail only appears above 1 keV,
even above 1.2 keV for Vela, this excess will have some impact on the temperature 
measurement:
spectral fits with composite BB spectra will give slightly 
lower $T_e^\infty$'s than single BB spectra and require lower $N_H$'s.
(Since the purely dipole case gives similar results we must correct the corresponding
statement made in Paper~I about spectral fits and replace it by the preceding one).
We will not attempt to quantify this small effect here since we restrict ourselves
to general characteristics of our model.
This excess in the Wien tail will also affect parameter values for the fits of the
hard tail for PSR 0656+14 and 1055-52 while in the case of Geminga our illustrative
fit of \S~\ref{sec:caps} can be done with the same parameter as used by Halpern \&
Ruderman (1993) for uniform surface temperature.

\section{RELIABILITY OF THE SURFACE TEMPERATURE MODEL 
            \label{sec:reliability}}

Our surface temperature model has two ingredients: 
\newline
1) the value of $T_s$ for a given field strength in the two cases of 
parallel and orthogonal transport [$T_s(\Theta_B =  0^\circ)$, and 
$T_s(\Theta_B = 90^\circ)$, respectively, where $\Theta_B$ is the 
angle between the local magnetic field and the radial direction], 
which are given in figure 1 of Paper I for various field strengths 
$B$, and 
\newline
2) the dependence of $T_s$ on $\Theta_B$ for a given field strength $B$:
\be
T_s(\Theta_B) = \chi(\Theta_B) \times T_s(\Theta_B = 0^\circ)
\label{equ:Ts1}
\ee
where
\be
\chi(\Theta_B) = (\cos^2 \Theta_B + \chi_0^2 \sin^2 \Theta_B )^{1/4}
         \;\; ; \;\; 
\chi_0 \equiv \frac{T_s(\Theta_b = 90^\circ)}{T_s(\Theta_b = 0^\circ)} \; .
\label{equ:chi}
\ee

As mentioned in Paper I, the dependence given in point (2) is a very
good approximation, much better than point (1), and its accuracy is
certainly better than the statistical uncertainties present in the
{\em ROSAT} data.
Most of the uncertainty of the model resides in the NS envelope calculations
used for point (1).
The `$T_b - T_s$ relationship' $T_s(T_b, \Theta_B,B)$, where $T_b$ is the 
interior temperature at the bottom of the envelope at density $\rho~=~10^{10}$~gm~cm$^{-3}$,
is known reasonably well 
at surface temperatures above $ 3 \cdot 10^5$ K in the case of 
$\Theta_B = 0^\circ$ and has been calculated by several authors who
obtained values in reasonable agreement with each other;
the problem is for the case $\Theta_B = 90^\circ$, where there are
large uncertainties due to several factors.
$T_s (\Theta_B = 90^\circ)$ is much lower than $T_s (\Theta_B = 0^\circ)$,
for a given $T_b$, 
and in this case of orthogonal transport the envelope structure is often
in the ``100\% Murphy regime'' where everything that can go wrong 
does go wrong:
electron conductivity perpendicular to the field is strongly
suppressed and photon transport is the dominant process up to much 
higher densities than in the case of parallel transport;
at these densities and low surface temperatures, the plasma temperature
is much higher than the local temperature but, thanks to the magnetic
field, extraordinary photons can still exist and transport heat;
lattice or ion heat transport can also contribute, a phenomenon that has 
been observed in laboratory materials under strong magnetic fields 
(Zyman 1960);
the low density region ($\rho < 10^{4-5} \rm g/cm^3$) can affect the surface
temperature and the magnetic effects on the equation of state become very 
large.
Finally, meridional flow of matter induced by the temperature gradients,
rotation, and magnetic field gradients can also smooth the temperature 
distribution.
Unfortunately, none of these effects has been taken into account in the
envelope calculation that we use for our model (Schaaf 1990), the only
such calculation published to date.

Despite of these limitations we consider our model as quite reliable thanks to
the results of \S~\ref{sec:quadr} which showed that the observable
characteristics of thermal emission do not depend sensitively on the
field strength for strengths above a few times 10$^{11}$ G
(when the emission spectra used are from blackbodies).
Maximum magnetic effects are obtained when $B$ is superior to about 
$10^{12}$ G, and saturate at higher fields.
For a large enough field $\chi_0 \ll 1$ and we can thus write
\be 
T_s(\Theta_B) = T_s(\Theta_B = 0^\circ) \times \cos^{1/2} \Theta_B
\label{equ:Ts2}
\ee
to a good approximation as long as $\Theta_B$ is not too close to
$90^\circ$; when $\Theta_B$ is close to $90^\circ$ the region is so
cold that it is not seen anyway.
The surface fields of the young pulsars we consider can be reasonably 
assumed to be above this threshold value of 2~--~3~$\cdot~10^{11}$~G:
values obtained by the standard magnetic dipole radiation braking
formula, as reported in table 1 of Paper I, are between $2.2 \cdot
10^{12}$ G (PSR 1055-52) and $8.8 \cdot 10^{12}$ G (PSR 0656+14)
without GR effects, and are even larger when GR effects are included 
(see above).
In the case of Geminga the optical data suggest the presence of a
synchrotron emission line which implies a magnetic field of at
least $3~\cdot~10^{11}$~G (Bignami {\em et al.} 1996).
So, as long as the exact value of $T_s (\Theta_B = 90^\circ)$ is about
3 times lower than $T_s (\Theta_B = 0^\circ)$, as it is at
$3~\cdot~10^{11}$~G in our model, for the field strengths actually present,
our interpretation of the {\em ROSAT} observations will remain valid.

\section{GRAVITATIONAL LENSING AND THE NS SIZE 
            \label{sec:lens}}

It was shown in Paper I that when only dipolar fields are used,
the pulsed fraction $Pf$ obtained is lower than what has been detected. 
The addition of a quadrupolar component can substantially increase
$Pf$ to values comparable to the observed ones as shown in \S~\ref{subsec:stat}
(figure~\ref{fig:stat}).
However, for this to happen we must choose carefully the strengths of the
various $\bf b_i$ compared to the dipole's strength and orientation:
random quadrupolar components generally give temperature distributions which are
so complicated that they flatten the light curves.

Vela and PSR 1055-52 have pulsed fractions, in the channel band where surface
thermal emission is detected, slightly higher than 10\%
and are considered to be almost orthogonal rotators:
the models of figure~\ref{fig:stat} with $\alpha \sim \zeta = 90^\circ$ thus apply
to them.
Taking into account that the effective temperature of  PSR 1055-52 is lower than 10$^6$~K
the observables $Pf$s would be slightly higher than values given in figure~\ref{fig:stat}.
We see that for 1.4 \Msol stars with radii of 13 or 10~km many models can produce
$Pf$'s as high as observed.
However only very few, special, field configurations can reproduce the
observed $Pf$ at $R = 7$ km, less than 0.1\% of all configurations generated
randomly for figure~\ref{fig:stat}.
The high $Pf~\sim$~20~--~30\% of Geminga, also an orthogonal rotator, is
however not reproducible with a 7~km radius 1.4~\Msol star 
(i.e., with $R = 1.7 R_S$):
the most strongly pulsed dipole+quadrupole model we found gives, at $T_e~=~5~\cdot~10^5$~K,
$Pf~\sim~15\%$ in channel band 7~--~50 as we have checked explicitly, and only a few models
at 10~km radius reach 20\% modulation.
In the case of PSR 0656+14, which has a BB effective temperature of
$\sim~8~\cdot~10^5$~K and $\alpha~\sim~\zeta~\sim~30^\circ$, we can obtain the
observed $Pf~\sim~14$\% with a 13~km radius 1.4 \Msol star with many models
and at 10~km radius in only a few cases, but with no model at all for a 7~km star:
in this latter case the highest $Pf$ we can obtain is about 4\%.
If we take $\alpha \sim \zeta \sim 8^\circ$ (Lyne \& Manchester 1988) then
$Pf$ is always extremely small:
even with a radius of 13~km, for a mass of 1.4~\Msol, the same procedure as used
for figure~\ref{fig:stat} could not find any model which gives a $Pf$ higher that
4\%.

\section{ENERGY AND TEMPERATURE DEPENDENCE OF THE PULSED FRACTION
            \label{sec:pf}}

One of the most important characteristics of the observed pulsations is
the energy dependence of the pulsed fraction $Pf$.
We showed in Paper I that BB emission with
dipolar fields always produces an {\em increase} of $Pf$ with photon energy.
We confirm this here, with a few exceptions, for more general
temperature distributions and explain the origin of this feature.
BB emission, with the appropriate surface temperature distribution,
can explain many of the observed characteristics of surface thermal emission of
the four pulsars we are studying.
Since there are many theoretical reasons for BB emission to be inadequate
it is important to find at least one property not reproducible in our simple
model in order to guide us toward the correct atmosphere model(s).
The energy dependence of $Pf$ happens to be such a feature:
as mentioned in \S~\ref{sec:data} in the three cases of PSR 0656+14, Geminga,
and PSR 1055-52, $Pf$ decreases at channels around channel 40~--~50.
Another strong constraint on atmosphere models is of course that the spectral
fit must imply a reasonable pulsar distance (cf. discussion in \S~\ref{sec:concl}).

The reason for the increase of $Pf$ with energy for BB emission
was elucidated by Page, Shibanov, \& Zavlin (1995):
the hardness of the BB spectrum {\em increases} with energy.
The ratio of the BB fluxes emitted at energy $E$ by two regions of 
areas $A_1$ and $A_2$, with temperatures $T_1$ and $T_2$ respectively, is
\be
\frac{F_{BB}(E,T_1)}{F_{BB}(E,T_2)} = \frac{A_1}{A_2} \times 
   \frac{\exp{(E/k_B T_2)}-1}{\exp{(E/k_B T_1)}-1}
\ee
which is an {\em increasing} function of $E$ if $T_1 > T_2$. 
This means that if we have a warm spot at temperature $T_1$ on a
surface with uniform temperature $T_2 < T_1$, then $Pf$ naturally
increases with energy.
A smooth temperature distribution with only one peak in temperature
will obviously give the same result.
With several peaks, by experience we can state that $Pf$ always increases
(with increasing channel energy) up to channels $\sim$~60 for $T_e~\sim~5~\cdot~10^5$~K
and up to channel $\sim$~100 for $T_e~\sim~10^6$~K:
we have tried {\em many} models with dipoles, off-centered dipoles,
quadrupoles, dipole+quadrupoles, warm plate(s) and found no exception.
Only in a few special cases, we have `observed' a decrease in $Pf$ in the
Wien tail which is due to the following (figure~\ref{fig:Pf-down}).
The surface presents two warm regions on opposite sides of the star, a very
large one and a smaller one where the temperature reaches a slightly higher
value (figure~\ref{fig:Pf-down}A), and only one peak is visible, i.~e., the
peak from the small warmer region does not appear as a peak but is only
filling the dip between the successive peaks of the large region
(figure~\ref{fig:Pf-down}B); by looking at increasing energy the filling of
the small peak increases (the small region is warmer than
the large one) and $Pf$ thus decreases (figure~\ref{fig:Pf-down}C).
If the area of the small warmer region is increased then it produces a
distinctive peak, i.~e., we obtain a double peak light curve, the
amplitude of both peaks increases with energy (the amplitude of the small
region peak will eventually win over the large region one) and $Pf$
increases.
It seems thus that the only mechanism, with BB emission, which can 
produce a decrease in $Pf$ is this filling of the interpeak by emission
from a small warmer region.
This is in fact the mechanism that Halpern \& Ruderman (1993) invoked
to explain the observed decrease of $Pf$ (in channel range 28~--~53
vs. 7~--~28) in Geminga but with an essential difference:
they proposed that the hard tail component, which is about $100^\circ$
off phase with the thermal component, would fill the dip in the light curve.
However the hard tail is much harder than the surface thermal component,
i.~e., clearly separable from it in the spectrum, and cannot
actually produce a decrease in $Pf$ at energies below 0.5 keV as we show
in the next section.
In the cases where we have `observed' a decrease in $Pf$, the
responsible component is only slightly warmer than the rest of the
star and its emission is practically indistinguishable in the
spectrum.

Another feature encountered in Paper I for dipolar fields was a
steepening of the increase of $Pf$ in the channel band 50~--~70 which
is due to the response of the PSPC.
The carbon present in the detector's window produces a strong absorption
edge at 284~eV: the effective area $A(E)$ vanishes just above this energy
and starts growing significantly only above 400~eV.
Thus, channels below $\sim$~40 pick up almost exclusively photons with energy
below 284~eV while channels above $\sim~50$ detect photons almost exclusively
above 500~eV: 
since the `absolute' pulsed fraction, with BB emission, is growing with energy
there is naturally a sharp rise of the observed $Pf$ in the channel
band 50~--~70.
This feature must hence be present in any model producing a rise of $Pf$ with energy
and can be seen, for example, in figures~\ref{fig:Pf-down}C and \ref{fig:Geminga}C.

For a given field configuration, the observable $Pf$ at a given energy 
increases when the overall stellar effective temperature decreases.
We stated, wrongly, in Paper I that this is due to the increase of the
temperature difference between the warm and cold regions when $T_b$
decreases.
However, a much stronger temperature difference increase is induced by the
increase of the field strength and it has almost no effect on $Pf$
as stated in \S~\ref{sec:strength}.
The actual reason for the increase of $Pf$, at a given energy, with
decreasing $T_e$ is the decrease of $Pf$ with photon energy since
lowering $T_e$ pushes the Rayleigh-Jeans part of the spectrum (which has
a low $Pf$) out of the PSPC range.
A simple check of this statement has been done by considering `plate models'
with a uniform surface temperature $T_0$ on which two plates are superposed
at temperatures $T_{+}$ and $T_{-}$, respectively higher and lower than $T_0$:
keeping the ratios $T_+ : T_0 : T_-$ constant but lowering $T_0$ does
induce an increase in the observable $Pf$ (at a given energy).
If we consider the absolute $Pf$, i.e., as would be detected by a detector
with absolute energy resolution, then the absolute-$Pf$ vs.  $E$ curve (see
for example figure~\ref{fig:Pf-down}C or figure 8 in Paper I) is simply
shifted leftwards when $T_0$ is lowered, at constant $T_+ : T_0 : T_-$ ratios,
and, since $Pf$ increases with energy, at a given energy $Pf$ increases when
$T_0$ decreases.

\section{SEPARATION OF THE SURFACE THERMAL EMISSION FROM THE HIGH ENERGY TAIL
         \label{sec:caps}}

Our study so far, and in Paper I, has concentrated exclusively on the
soft X-ray band assuming implicitly that this thermal component can be
separated from the hard tail.
We argue here that this separation is justified, and show it explicitly
in the case of Geminga.
A simple look at the spectral fits (e.g., figure 9A below, or Halpern \& Ruderman 1993,
for Geminga; \"Ogelman 1995a for PSRs 0656+14 and 1055-42) shows
that, at energies slightly below the crossover of the soft thermal component
with the hard tail, the contribution of the latter becomes rapidly negligible
compared to the former.
Moreover, the spectral fit of the soft component is not changed significantly 
if the hard tail is fitted by BB emission or power law emission
(e.g., Halpern \& Ruderman 1993).
One thus does not expect much interference between these two components.

The contribution of the hard tail to the soft band is larger if it is modeled
as a power law rather than a BB.
To consider the worst case of interference between the two spectral
components we thus model Geminga's X-rays with a composite model
including surface thermal emission and a pulsed power law tail,
the results being shown in figure~\ref{fig:Geminga}.
The same surface temperature model complemented with a hard tail as thermal
emission from two polar caps has been presented in Page \& Sarmiento (1996).
The spectrum (figure~\ref{fig:Geminga}A) shows that the hard tail emission
is about 30 times weaker than the surface emission at channels below 40 but,
due to its strong pulsations, its contribution to the pulse profiles can
be about 10\% in the soft band 7~--~28 (figure~\ref{fig:Geminga}B1).
Nevertheless this is far from enough to alter the shape of the light curves
in both bands 7~--~28 and 28~--~53 (figure~\ref{fig:Geminga}B1 \& B2).
The important consequence of this is the effect on the energy dependence of
the pulsed fraction shown in figure~\ref{fig:Geminga}C: 
the general trend, typical of BB emission, consisting of an increase of $Pf$ with
energy (see \S~\ref{sec:pf}) is slightly weakened by the addition of the
hard tail but not reversed at channels below 100 (figure~\ref{fig:Geminga}C).
When modeled as polar cap thermal emission, the hard tail contribution
to the soft band is almost completely negligible (Page \& Sarmiento 1996).
It is thus not possible, within the present framework, to make the hard
tail responsible for the observed decrease of the pulsed fraction with
energy below 0.5~keV.
In conclusion, the observed decrease of $Pf$ with energy, at channels
below 53, observed in Geminga is most certainly an intrinsic property
of the surface thermal emission which is absolutely irreproducible with
BB emission.

The cases of PSR 0656+14 and 1055-52 are more delicate but, from a look
at the spectral fits there is no doubt that the observed decrease of
$Pf$ at channels below 70 and 50 respectively (\"Ogelman 1995a)
is also a property of the surface thermal emission since the crossover
with the hard tail occurs at channels about 100 for these two pulsars.
However the shift in the peak phase (\"Ogelman 1995a) which occurs
within the band dominated by the surface thermal emission is another
feature totally unexplainable with BB emission.
Thus, both the pulsed fraction variation and the peak phase shift
must be attributed to the anisotropy induced by the magnetic field
in the atmosphere, i.e., anisotropy of the emitted flux, superposed
to a non uniform surface temperature distribution.
In the case of Geminga the shift in the peak phase coincides with
the shift from the surface thermal emission to the hard tail and
is not surprising (\"Ogelman 1995b).

\section{SURFACE BOUNDARY CONDITION FOR NEUTRON STAR COOLING 
         \label{sec:cooling}}

A direct application of our surface temperature model and of the previous
analysis concerns the modeling of the neutron star thermal evolution.
In such models, the relationship $T_b - T_s$ between the temperature at the
bottom of the envelope $T_b$ and at the stellar surface $T_s$ is needed as an outer
boundary condition.
Taking into account the surface temperature inhomogeneity we must rather
speak of a $T_b - T_e$ relationship, where $T_e$ is the effective temperature
of the whole surface, and, given different field structures,
we can easily generate many such $T_b - T_e$ relationships.
Cooling models including magnetic effects in the envelope have been presented 
by Van Riper (1991) and Haensel \& Gnedin (1994) who used a radial field 
$T_b~-~T_e$ relationship:
this is not a realistic configuration and we will show here
that a more careful treatment actually leads to conclusions opposite to the
one reached by these authors.
For a radial field, $T_e$ is increased compared to the non magnetic case,
at a fixed $T_b$, as shown in figure 1 of Paper I:
during the photon cooling era, i.~e., neutron star age above $\sim
10^5$ yrs, this results in an increased surface photon emission, for a
given internal temperature $T_b$, and accelerated cooling compared to 
the non magnetic case.
For example, with a radial field of strength $10^{13}$~G, $T_e$ is
increased by about 25\%, compared to the non magnetic case, and thus
the photon luminosity is increased by a factor of 2.5.

The minimal inclusion of the magnetic field in the $T_b - T_e$ relationship
should be done with a dipolar geometry.
By integrating Eq.~\ref{equ:Ts1} for a dipolar field one easily obtains
\be
T_e (T_b) = T_s(T_b,\Theta_B = 0) \cdot [1 - 0.47 (1-\chi_0^{\; 4})]^{1/4}
\label{equ:te-ts}
\ee
for the effective temperature $T_e$ as a function of the maximum surface temperature
$T_s(\Theta_B = 0)$ and $\chi_0$ (Eq.~\ref{equ:chi}):
this reduces the magnetic field effects and brings the $T_b - T_e$ relationship
close to the non magnetic one as is illustrated in figure~\ref{fig:tbte}.
Since we have shown that the surface field of the four pulsars which show
surface thermal emission (and can be compared with cooling models) is
not dipolar, we must then include the quadrupolar component in the $T_b - T_e$
relationship.
A general idea of the effect can be immediately seen from figure~\ref{fig:stat}
which plots \mbox{$T_e$(mag.)/$T_e$(non mag.)}.
At $T_b$ = 10$^8$ K, i.~e., $T_e \sim 10^6$ K, most configurations also
induce a {\em decrease} of $T_e$ compared to the non magnetic case.
Moreover, \mbox{$T_e$(mag.)/$T_e$(non mag.)} decreases for configurations with
larger $Pf$ and is systematically smaller than one for the cases
which can produce the large observed pulsed fractions.
We also show in figure~\ref{fig:tbte} a second $T_b - T_e$ relationship
which corresponds to the most strongly pulsed configuration that we have 
found with a dipole+quadrupole field.
The interesting result is that this relationship does not depend significantly
on the field configuration: it is very similar to the dipolar case.
We have explicitly verified that other field configurations which
produce large observable pulsations give almost identical results,
as can also be seen from figure~\ref{fig:stat} (for $T_b = 10^8$ K).
The discussion at the end of \S~\ref{subsec:stat} anticipated this result
and explained it beforehand.
Configurations of figure~\ref{fig:stat} which give low $Pf$'s can however
increase $T_e$ by 10\% compared to the non magnetic case:
the observed strong pulsations, if really due to
large surface temperature inhomogeneities, totally rule out such field
configurations and thus such boundary conditions.
From this we conclude that {\em the most likely overall effect of the
magnetic field, in realistic configurations, is at most a slight
decrease of the heat flow in the envelope as compared to the non magnetic case.}
Notice that the 10$^{13}$ G case is practically identical to the non magnetic case,
a result already obtained by Hernquist (1985).
Since the strongly pulsed dipole+quadrupole field configurations give $T_b - T_e$
relationships almost identical to the purely dipolar case, one could use
the dipolar relationship for magnetic cooling models.
The dipolar strengths usually quoted are obtained from the standard
magnetic dipolar radiation braking formula and thus only indicative of
the actual surface field, and even if that formula were exact, GR
effects can increase the surface field as shown
in figure~\ref{fig:corr}.
We thus propose to take as a boundary condition for neutron star cooling models
\be
T_e = a \; g_{s 14}^{1/4} \; T_{b 8}^b,
\label{equ:tbte}
\ee
where $a = 0.85^{+0.05}_{-0.20}$ and $b \cong 0.55$.
The value of $a = 0.85$ is an average of the result of Hernquist (1985) who obtained 
0.83 and Gundmundsson, Pethick, \& Epstein (1982) who obtained 0.87 and the uncertainties
take into account the magnetic field effect:
+0.05 corresponds to $B \sim 10^{13}$~G and -0.20 to $B \sim 10^{11}$~G.
The exponent $b = 0.55$ is from figure~\ref{fig:tbte} and corresponds also to the value of
Gundmundsson {\em et al.} (1982).

The cooling calculations of Page (1994), restricted to the photon
cooling era where the $T_b - T_e$ relationship is most important,
where explicitly performed with the non magnetic case in anticipation
of the present results and his conclusions about the necessity of
extensive baryon pairing in the core of the Geminga neutron star are
therefore still valid.
For illustration, we show in figure~\ref{fig:cool} several cooling
curves with four boundary conditions from figure~\ref{fig:tbte} which show
explicitly the smallness of the magnetic field effects.

When discussing pulse profiles the reliability of the envelope models
is not very important as discussed in \S~\ref{sec:reliability} since most
of the effect simply comes from the $\cos^{1/2} \Theta_B$ geometrical factor in
equation~\ref{equ:Ts2}.
When considering the $T_b - T_e$ relationship the reliability of envelope models
for $\Theta_B = 90^\circ$ is also not very important since the cold regions
are so cold that they contribute very little to the overall effective temperature,
[as shown explicitly by the $(1-\chi_0^{\; 4})$ dependence of $T_e$ for dipolar fields
in eq.~\ref{equ:te-ts}].
Envelope models for $\Theta_B = 0^\circ$ are what really determines the $T_b - T_e$
relationship and more reliable calculations at low temperature are needed for
better interpretation of cool neutron star observations which will most certainly
be detected with the future generations of X-ray satellites.

\section{DISCUSSION AND CONCLUSIONS 
         \label{sec:concl}}

\subsection{Modeling pulsars' surface temperature distribution}

We have completed our general study of thermal emission from magnetized neutron 
stars, within the framework of blackbody (BB) emission.
Our original intent was to determine if the inhomogeneous surface temperature
distribution induced by the anisotropy of heat flow in the neutron star
envelope is strong enough to produce the pulsed fraction observed
by {\em ROSAT} in Vela, Geminga, PSR 0656+14, and PSR 1055-52.
We have shown explicitly in \S~\ref{sec:reliability} that most of the
effect is in the geometrical factor $\cos^{1/2} \Theta_B$ (Eq.~\ref{equ:Ts2}),
as presented originally by Greenstein \& Hartke (1983), and fortunately
the actual value of $\chi_0$ (Eq.~\ref{equ:chi}) is of little importance.
Moreover, models of magnetized neutron star envelopes allow us to relate the interior
temperature $T_b$ to the maximum surface temperature $T_s(\Theta_B~=~0^\circ)$
quite reliably as long as the latter is higher than about $3~\cdot~10^5$~K.
The minimum surface temperature $T_s(\Theta_B~=~90^\circ)$ is however poorly
determined but its actual value, equivalent to $\chi_0$, is not really important.
In short, {\em the surface temperature distribution of a magnetized neutron star
can be modeled well enough for present practical purposes}.

Since we do not take into account the effect of the magnetic field on the
emitted spectrum but only on the local surface temperature, our results
are not sensitive to the actual strength of the field.
As long as the magnetic field is stronger than a few times 10$^{11}$~G
its effects saturate and we are mostly dealing with warm plates whose
size is determined by the field orientation $\Theta_B$.
When atmospheric effects are included one can expect a real dependence
on the field strength due to the presence of absorption edge(s) and a field
dependent anisotropy of the emission.
As in the case of dipolar fields (Paper I), the composite BB spectrum produced
by the temperature nonuniformity is close to a BB at the star's effective 
temperature with some excess in the Wien tail (figure~\ref{fig:comp-spectrum}).
The quadrupole+dipole configurations which induce strong pulsations show an
excess similar to the pure dipolar case since they have warm and cold regions 
of similar areas but many of the configurations which produce little pulsations
give a composite BB spectrum closer to a single temperature BB spectrum since
they have many warm regions and smaller cold regions.

\subsection{The quadrupolar component}

We showed in Paper I that dipolar surface fields do not allow to reproduce the
amplitude of the observed pulse profiles;
however, we have shown here that the inclusion of a quadrupole component
is sufficient to raise the pulsed fraction up to or above the observed values.
Gravitational lensing is crucial in this result, and the effect of the
quadrupole, in the cases where it does increase the pulsed fraction, is
basically to push the two magnetic poles, and the warm regions surrounding
them, closer to each other: 
the areas of the warm and cold regions do not change substantially, compared
to the purely dipolar case, but gravitational lensing is then less effective
in keeping constantly a warm region in sight.
This shift of the magnetic poles location is in fact immediately seen
in the cases of Geminga and PSR 1055-52 which are considered to be orthogonal
rotators but show only a single wide peak in the soft X-ray band
while a dipolar field would produce two peaks
(see also discussion in Halpern \& Ruderman 1993).
If we want to stick to a multipolar description, strong octupole
components would make the matter worse, inducing still more
warm region and flattening even more the light curves.
Thus, the simple observation of significant pulsations imply that the
surface field is dominated by the dipole with a significant quadrupole
component, but nevertheless weaker than the dipole, and that higher
order components must be weaker.
It has been proposed that the presence of the quadrupole is also felt
in the anomalous dispersion measure of the high frequency vs. low
frequency radio pulses which has been observed in several pulsars
(Davies {\em et al.} 1984; Kuz'min 1992).
{\em Similar observations of PSR 0833-45, 0656+14, and 1055-52, would allow
a direct comparison with the strengths of the quadrupoles needed for the 
interpretation of the radio and X-ray emission of these pulsars.}
No confirmation of the presence of a strong quadrupole component
could help constraining the size of these neutron stars through gravitational
lensing.
Another possibility is that the multipole expansion is inappropriate to
describe the surface field (i.~e., would require the inclusion of many
strong high order components) as would be the case if the field has a
sunspot like structure:
plate tectonics (Ruderman 1991a, b, c) does predict such a structure
which may be appropriate for cases like Geminga (Page {\em et al.} 1995).

Most dipole+quadrupole configurations produce little
observable pulsations and it is surprising that all four observed
pulsars have chosen an {\em a priori}
highly improbable configuration which does increase $Pf$.
If we visualize the quadrupole as a pair of coplanar dipoles of equal
strengths but opposite directions, then the dipole+quadrupoles configurations 
which are successful in increasing $Pf$ are almost coplanar: 
there must be a good physical reason for this to happen in all four cases.
Possible identification of a new young neutron star in a supernova remnant close to
CTB 1 has been recently reported (Hailey \& Craig 1996): 
this object appears to be similar in age and temperature to the Vela pulsar and
also shows pulsations at a 10\% $\pm$ 10\% level.
If the detected X-rays are from the surface thermal emission and $Pf$ is above 10\%
this would raise to five the number of neutron stars in which a quadrupole with
adequate orientation is present.
The theory of plate tectonics (Ruderman 1991a, b, c) predicts that the crust of
young fast spinning pulsars should break under the stress produced by the internal
superfluids and superconductors.
Plate motion, towards the equator, will ensue and the mutual attraction of the two 
magnetic poles may slowly pull them toward each other (as is apparently seen in
Geminga and PSR 1055-52) inducing a strong quadrupole component as we need in the
present study.

\subsection{Constraints on the neutron star size.}

We found that, within our present study, the observed pulsed fractions of
Geminga and PSR 0656+14 cannot be reproduced if these stars have a radius of
7~km for a mass of 1.4 \Msol, or, more generally, if their radius is of the order
of $1.7~\cdot~R_S$, $R_S$ being the star's Schwarzschild radius.
Moreover, a 10~km radius for a 1.4 \Msol star, i.e., $R~=~2.4~R_S$, can be
`rejected at the 99\% confidence level' in the sense that less than 10 models
out of 1000 generated for figure~\ref{fig:stat} can reproduce the observed $Pf$
for these same two neutron stars.
Similarly, for Vela and PSR 1055-52, $R~\sim~1.7~R_S$ can be also be `rejected
at the 99\% confidence level' for the same reason but $R~=2.4~R_S$
is acceptable.
Our results thus favor rather large radii, above 10~km for a 1.4 \Msol mass,
for all four neutron stars.
These conclusions must be taken with extreme caution for several reasons:
1) the dipole + quadrupole description of the surface field may not be adequate,
2) anisotropic emission due to the magnetic field may increase significantly the
observable pulsed fraction,
3) the emitted flux may be partially absorbed in the magnetosphere,
4) the surface temperature may be controlled by factors other than heat flow from the
hot interior.
Each of these factors may, however, increase or reduce the observable pulsed fraction.
Thus, we consider our results as mostly illustrative but notice that it is
encouraging that the observed pulsed fractions fall within the range where we
could potentially rule out large classes of neutron star models through gravitational
lensing of the surface thermal emission.
We finally mention that gravitational lensing of the hot polar cap emission may
also provide complementary estimates of the neutron star radius:
Yancopoulos, Hamilton, \& Helfand (1994) find $R~=~(2.4~\pm~0.7)~R_S$, or
$10~\pm~3$~km at 1.4~\Msol, for PSR 1929+10 but the anisotropy of emission and
the lack of information on the location of the second polar cap make this estimate
also uncertain.

\subsection{Limitations of blackbody emission}

We confirmed our previous result that with BB emission the observable pulsed
fraction $Pf$ is unavoidably increasing with photon energy, at low energy
where the surface thermal emission is detected, and explained that this is due
to the hardening of the BB spectrum with increasing
temperature.
The three pulsars Geminga, 0656+14, and 1055-52 show a decrease of $Pf$ with
energy in the soft band, irreproducible with BB emission.
The model of Page {\em et al.} (1995) produced such a decrease of $Pf$ by
superposing two different spectra, a non magnetic hydrogen atmosphere at low
temperature on most of the stellar surface and a magnetized hydrogen atmosphere
on two warm magnetized plates: the magnetic spectrum is
much softer than the non magnetic one and $Pf$ decreases strongly with energy.
Extrapolating from this result we may speculate that the decrease of $Pf$ with energy is
due to a softening of the emitted spectrum with increasing temperature.
We also showed in \S~\ref{sec:caps} that the separation of the soft thermal
component and the hard tail is large enough to make the contribution of the latter
in the soft band quite small:
the observed decrease of $Pf$ is an intrinsic feature of the surface thermal
emission and not due to an interference of these two components as proposed
by Halpern \& Ruderman (1993) in the case of Geminga.

A second clear indication for the inadequacy of BB, or BB-like, emission is
directly found in the spectral fits: for a reasonable neutron star size the required
distance $D$ must agree with other estimates as, e.~g., the distance obtained
from the pulsar's dispersion measure or from parallax measurements (Geminga).
However, the distance $D$ and stellar radius $R^\infty$ obtained from
spectral fits have to be taken with caution due to the uncertainty in
the calibration of the PSPC in the lower channels band;
the fit in this band determines the column density $N_H$ which
strongly affects the estimate of the emitted flux and the PSPC
calibration uncertainty translates into uncertainty on $D$ and
$R^\infty$ (Meyer, Pavlov, \& M\'{e}sz\'{a}ros 1994).
In the case of Vela, BB emission is disastrous since it implies a 3~-~4~km 
radius neutron star at 500~pc or, alternatively, a distance of
$\sim$~1500~pc for a standard size neutron star (\"Ogelman {\em et al.} 1993).
The same is apparently happening in the case of the newly discovered
neutron star RXJ 0002+6246 (Hailey \& Craig 1996) for which BB spectral fits also
require an effective blackbody radius $R^{\infty}~\sim~2.5~-~4.5$~km for the
estimated distance of 3~kpc.
However, magnetized hydrogen atmosphere models do pass the distance test and can
fit the Vela spectrum with the standard distance of 500~$\pm$~125~pc 
(Page, Shibanov, \& Zavlin 1996) and a reasonable hydrogen column density.
In the case of Geminga, with a parallax measured distance of $157^{+59}_{-34}$~pc
(Caraveo {\em et al.} 1995), available magnetized hydrogen atmosphere models do
pretty bad since they need a distance of about 20~pc (Meyer {\em et al.} 1994;
Page {\em et al.} 1995) while BB is acceptable with a
required distance between 150 and 400~pc (Halpern \& Ruderman 1993).
This low distance needed for magnetized hydrogen atmosphere spectral fits
may, however, be partly due to the uncertainties in the PSPC calibration at low 
energies (Meyer {\em et al.} 1994).
The crucial difference between Vela and Geminga is in the temperature:
at 10$^6$ K (Vela) the hydrogen atmosphere is fully ionized and the 
models are reliable while in Geminga, at $T~\sim~3~-~5~\cdot~10^5$~K, the
atmosphere is only partially ionized and the available atmosphere models are
not yet very accurate.
Atomic thermal motion (Pavlov \& M\'{e}sz\'{a}ros 1993) and its
effects on the bound-bound (Pavlov \& Potekhin 1995) and bound-free
(Pavlov \& Potekhin 1996) opacity will affect significantly the
absorption edge present at $T_e$ below 10$^6$~K, but inclusion of these
effects into atmosphere models has not yet been done and the results are hard to
predict.
The field and temperature dependence of the absorption edge may naturally
induce a decrease of $Pf$ with increasing photon energy as observed
(Pavlov 1995).

These limitations of BB emission may cast doubts on the validity of our
results.
As we stated in the introduction we consider that modeling NS thermal
emission with BB emission is a mandatory first step and our study
does allow us to delimitate more clearly the limitations of this approach.
BB is certainly not adequate for reliable estimates of pulsars' surface temperature,
as well as size and distance, from spectral fits.
However, modeling of the pulse profiles is probably less dependent on the exact
atmosphere characteristics since the anisotropy of the emission from the atmosphere
is not very strong at low energy and it is unlikely to affect dramatically our 
conclusions on the effect of gravitational lensing and neutron star sizes.
Our results obviously have to be taken only as a first step which hopefully
can indicate the direction to be taken for future work(s).

\subsection{Neutron star cooling}

We have finally applied our model to assess the effect of the crustal magnetic field
on the outer boundary condition used in neutron star cooling models.
We have shown that the global effect of the magnetic field, compared
to the non magnetic case, is very small and most probably consists
in a slight reduction of the heat flux in the envelope which results in a
lower effective temperature $T_e$ for a given inner temperature $T_b$.
This is in sharp distinction to the case where magnetic effects are
naively applied under the assumption of a uniformly magnetized
surface, implying an increase of $T_e$ as compared to the non magnetic
case.

\subsection{Warning about the relevance of the model}

Finally, we must repeat the warning expressed in Paper I about the relevance of
our study:
strong heating of the surface by magnetospheric hard X-rays and/or substantial
magnetospheric absorption of the emitted flux, as proposed by Halpern \& Ruderman 
(1993), could invalidate seriously our results (Page 1995b).
The possible detection of pulsed $\gamma$-rays from PSR 0656+14 (Ramanamurthy 1995)
would mean that all four pulsars we study are copious $\gamma$-ray emissors
and heating of the neutron star surface by such energetic magnetospheres is
not an unreasonable hypothesis.
In the original Halpern \& Ruderman (1993) model for Geminga the flow of
electrons/positrons onto the polar caps produced a luminosity
$L_p~\sim~2.6~\cdot~10^{32}$~ergs s$^{-1}$, radiated away as hard X-rays
which would then be back scattered onto the neutron star
surface and reemitted as soft surface thermal X-rays.
The X-ray luminosity of Geminga is about $2~\cdot~10^{31}$~ergs~s$^{-1}$ for
the soft thermal component and $8~\cdot~10^{29}$~ergs~s$^{-1}$ for the hard
tail (for a distance of 160 pc as measured by Caraveo {\em et al.} 1995) so
that even a small proportion of the hard polar X-rays hitting back the surface
could
explain the observed temperature.
The problem of this model was that the {\em total} predicted X-ray luminosity is
more than one order of magnitude higher than observed.
The argument has been recently revised and the polar cap luminosity
scaled down by Zhu \& Ruderman (1996) who now obtain 
$L_p~\sim 8~\cdot~10^{30}$~ergs~s$^{-1}$.
This is much closer to what is observed and may be the factor controlling
Geminga's surface temperature.
However, the soft X-ray flux is about 30 times larger than the hard one
so that in this model almost all the polar cap hard X-rays must be
scattered back onto the stellar surface.
On the other side, neutron star cooling models are also perfectly able to
explain the observed temperature (Page 1994).
We can only hope that future work(s) will elucidate this dilemma.

\acknowledgments 

We are grateful to Yu. A. Shibanov for many discussions and G.~G. Pavlov for
comments on part of the manuscript.
This work was supported by a UNAM-DGAPA grant (No. IN105794 continued through
grant No. IN105495)
D.~P. acknowledges support from a {\em C\'atedra Patrimonial} of
CoNaCyT.


\clearpage 
\begin{footnotesize}
\begin{deluxetable}{cccccc}
\tablewidth{0pt}
\tablecaption{\label{tab:quadr}
              Spherical components of the five generating quadrupolar fields and location of the magnetic
              poles.}
\tablehead{
      & $\bf b_0$ & $\bf b_1$ & $\bf b_2$ & $\bf b_3$ & $\bf b_4$}
\startdata 
$b_r $   &       $ \frac{3}{2} \cos^2 \theta - 1 $ &
                 $             \sin 2 \theta \; \sin \phi $ &
                 $           - \sin 2 \theta \; \cos \phi $ &
                 $           - \sin^2 \theta \; \sin 2 \phi $ &
                 $             \sin^2 \theta \; \cos 2 \phi $ \nl
$b_{\theta}  $ & $   \frac{1}{2} \sin 2 \theta $ &
                 $ - \frac{2}{3} \cos 2 \theta \; \sin \phi $ &
                 $   \frac{2}{3} \cos 2 \theta \; \cos \phi $ &
                 $   \frac{1}{3} \sin 2 \theta \; \sin 2 \phi $ &
                 $ - \frac{1}{3} \sin 2 \theta \; \cos 2 \phi $ \nl
$b_{\phi} $  &   $     0       $  &
                 $ - \frac{2}{3} \cos \theta \; \cos \phi $ &
                 $ - \frac{2}{3} \cos \theta \; \sin \phi $ &
                 $   \frac{2}{3} \sin \theta \cos 2 \phi $ &
                 $   \frac{2}{3} \sin \theta \sin 2 \phi $ \nl
`North' poles  &
     $\theta = 0 \; \& \; \pi$  &
         $( \theta , \phi ) = \left\{ \begin{array}{c} (\frac{\pi}{4},\frac{\pi}{2}) \\
                                               (\frac{3\pi}{4},\frac{3\pi}{2}) \end{array} \right.$ & 
           $( \theta , \phi ) = \left\{ \begin{array}{c} (\frac{\pi}{4}, \pi) \\
                                               (\frac{3\pi}{4},0) \end{array} \right.$ & 
             $( \theta , \phi ) = \left\{ \begin{array}{c} ( \frac{\pi}{2},\frac{3\pi}{4}) \\
                                               (\frac{\pi}{2},\frac{7\pi}{4}) \end{array} \right.$ & 
               $( \theta , \phi ) = \left\{ \begin{array}{c} (\frac{\pi}{2},0) \\
                                               (\frac{\pi}{2},\pi) \end{array} \right.$ \nl
`South' poles  &
     $\theta = \frac{\pi}{2}$  &
         $( \theta , \phi ) = \left\{ \begin{array}{c} (\frac{\pi}{4},\frac{3\pi}{2}) \\
                                               (\frac{3\pi}{4},\frac{\pi}{2}) \end{array} \right.$ & 
           $( \theta , \phi ) = \left\{ \begin{array}{c} (\frac{\pi}{4},0) \\
                                               (\frac{3\pi}{4},\pi) \end{array} \right.$ & 
             $( \theta , \phi ) = \left\{ \begin{array}{c} ( \frac{\pi}{2},\frac{\pi}{4}) \\
                                               (\frac{\pi}{2},\frac{5\pi}{4}) \end{array} \right.$ & 
               $( \theta , \phi ) = \left\{ \begin{array}{c} (\frac{\pi}{2},\frac{\pi}{2}) \\
                                               (\frac{\pi}{2}, \frac{3\pi}{2}) \end{array} \right.$ \nl
\enddata
\end{deluxetable}
\end{footnotesize}

\clearpage 

 
\clearpage

\begin{figure} 
\caption{Gravitational correction factors for the radial ($r$), f$_l$, and
         spherical ($\theta$,$\phi$), g$_l$, parts of the
         dipole ($l = 1$) and quadrupole ($l = 2$) components of
         the magnetic field at the stellar surface as a function of
         the ratio of the star's radius $R$ to its Schwarzschild
         radius $R_S$ or of the actual star's radius at masses
         of 1.2, 1.4, and 1.6 \Msol.
\label{fig:corr}}
\end{figure}

\begin{figure} 
\caption{Statistics of observable pulsations with dipole+quadrupole magnetic fields.
         For each stellar radius and dipole-observer orientation ($\alpha$-$\zeta$)
         with respect to the rotational axis 1000 surface magnetic field and the
         induced temperature distribution were generated
         The x axis plots the observable pulsed fraction in the PSPC channel band 7 -- 50
         for the given surface temperature distribution
         and the y axis the ratio of the star's effective temperature for the given magnetic
         field configuration to the temperature it would have with zero field.
         The surface temperature are calculated from an interior temperature $T_b = 10^8$~K: 
         all models have an effective temperature around 10$^6$~K. 
         Lowering $T_e$ does increase
         $Pf$: by about 15\% at $T_e \sim 3 \cdot 10^5$~K and 30\% at 10$^5$~K.
         See text, \protect\S~\protect\ref{subsec:stat}, for more detail.
\label{fig:stat}} 
\end{figure}

\begin{figure} 
\caption{Surface temperature distribution for the dipole + quadrupole magnetic field
         configuration of figure~\protect\ref{fig:stat} which gives the highest
         observable pulsed fraction.
         The whole star's surface is shown in an area preserving mapping.
         With an internal temperature of $10^8$~K, the effective
         temperature is $9.33~\cdot~10^5$~K while the maximum and minimum
         temperatures are, respectively, 
         $1.2~\cdot~10^6$~K and $3.1~\cdot~10^5$~K
         (1.4~\protect\Msol star with a 10~km radius).
         Four sets of isotherms are plotted as indicated.
         Field components:
         dipolar component of strength $10^{12}$ G and orientation
         $\theta = 90^\circ$ and $\phi = 0^\circ$, 
         and quadrupolar components of strengths
         Q$_0~=~2.48~\cdot~10^{10}$~G,
         Q$_1~=~6.84~\cdot~10^{10}$~G,
         Q$_2~=~5.01~\cdot~10^{10}$~G,
         Q$_3~=~4.80~\cdot~10^{11}$~G, and
         Q$_4~=~-6.24~\cdot~10^{10}$~G:
         the maximum field strength at the surface is $2.18~\cdot~10^{12}$~G
         and the minimum $1.72~\cdot~10^{11}$~G.
         The two black dots show the approximate locations of the polar caps, obtained
         by integrating the last open magnetic field lines from the 
         light cylinder down to the stellar surface while the two small circles at
         $\theta = 90^\circ$ and $\phi = 180^\circ$ and $360^\circ$ show their
         positions without the quadrupolar component.
\label{fig:model1-temp}}
\end{figure}

\begin{figure} 
\caption{Effect of the inhomogeneous surface temperature on the spectrum.
         A: incident flux (i.e. with redshift and lensing) at five temperatures,
         with zero interstellar absorption.
         B: the same fluxes with interstellar absorption and passed through the PSPC'
         response matrix.
         The temperatures of 5, 7, 9 and 15 $\cdot 10^5$ K correspond to the estimated
         blackbody temperatures of Geminga, PSR 1055-52, PSR 0656+14 and Vela 
         respectively.
         We used a 1.4 \Msol, 10 km radius ($R^\infty$ = 13.06 km) star, at a distance
         of 500 pc, with the same field configuration as in figure~\protect\ref{fig:model1-temp}.
         Continuous curves: composite BB spectra, dashed curves: single temperature BB spectra
         at $T = T_e$.
\label{fig:comp-spectrum}} 
\end{figure}

\begin{figure}
\caption{A model producing a decreasing pulsed fraction with blackbody emission.
         (1.4 \Msol star with a 12 km radius.)
         \newline
         {\bf A.} Surface temperature distribution. 
         Magnetic field: dipole orthogonal to 
         the rotation axis (in direction $\theta = 90^\circ$ and $\phi = 90^\circ$)
         with its center shifted by a half radius in direction of the north magnetic pole.
         The warm region around the north pole reaches a temperature of
         $8.15 \cdot 10^5$ K but is much smaller than the warm region 
         around the south pole whose maximum temperature is $7.25 \cdot 10^5$ K.
         \newline
         {\bf B.} Light curves in three channel bands produced by the temperature
         distribution of A, as seen through the {\em ROSAT}'s PSPC by an observer
         at 150 pc and $N_H = 10^{20} \; {\rm cm^{-2}}$.
         Star's rotation axis along $\theta = 0^\circ$ and observer at
         $\theta = 90^\circ$.
         The dotted lines show light curves where the temperature of the smaller
         north magnetic pole region has been blocked at $7.25 \cdot 10^5$ K, i.~e., the
         same maximum temperature as the large region around the south magnetic pole:
         this clearly shows that this warmest region is responsible for the decrease 
         of the $Pf$.
         \newline
         {\bf C.} Observable (with the PSPC) and absolute 
         (i.~e., as would be seen with a perfect energy
         resolution detector) $Pf$. The dotted curves show the corresponding $Pf$ 
         when the temperature of the small warm region has been blocked at
         $7.25 \cdot 10^5$ K (as in B).
\label{fig:Pf-down}}
\end{figure}

\begin{figure} 
\caption{Composite model of Geminga's soft X-ray emission:
         surface thermal emission (dashed lines) plus 
         power law emission (dotted lines)
         and total emission (continuous lines).
         Surface temperature distribution induced by a dipole+quadrupole
         magnetic field: dipole of strength $10^{12}$~G perpendicular to
         the rotational axis and quadrupole with components
         $Q_0~=~-2.28~\cdot~10^{11}$~G,
         $Q_1~=~+4.95~\cdot~10^{10}$~G,
         $Q_2~=~+3.36~\cdot~10^{10}$~G,
         $Q_3~=~+7.08~\cdot~10^{11}$~G,
         $Q_4~=~-4.53~\cdot~10^{11}$~G.
         The interior temperature is $4.04~\cdot~10^7$~K, giving an effective
         temperature at infinity $T_e^\infty~=~4.30~\cdot~10^5$~K for a
         1.4~\Msol star with a radius of 12~km.
         The hard tail is a sum of two components, both power laws with indices
         1.47, modulated with the rotational phase $\Phi$ by a $\cos~\Phi$ factor.
         The observer is located within the rotational equatorial plane at a distance
         of 185~pc and $N_H~=~2.1~\cdot~10^{20}$~cm$^{-2}$.
         A. Observable count rates from the surface thermal emission,
            the power law emission, and the total emission
            compared with the {\em ROSAT} data
            (from Halpern \& Ruderman 1993).
         B. Pulse profiles in three channel bands, compared with the {\em ROSAT} data
            (from Halpern \& Ruderman 1993). 
         C. Pulsed fraction for the two components and the total emission.
\label{fig:Geminga}} 
\end{figure}

\begin{figure} 
\caption{$T_b - T_e$ relationship: thick lines correspond to purely dipolar fields while
         the thin lines correspond to one of the maximally pulsed dipole+quadrupole configuration
         of figure~\protect\ref{fig:stat}. The indicated field strengths are the dipole component
         strength at the magnetic pole (without GR correction: so the actual field value is
         about 50 - 100\% higher, see figure~\protect\ref{fig:corr}).
\label{fig:tbte}} 
\end{figure}

\begin{figure} 
\caption{Effect of the crustal magnetic field on the cooling of neutron stars.
         We show the `standard' cooling of a 1.4 \Msun neutron star built with the FP
         equation of state (Friedman \& Pandharipande 1981) with and without
         core neutron $^3P_2$ pairing. 
         The neutron core pairing critical temperature $T_c$ is taken from Hoffberg 
         {\em et al.} (1970): the high $T_c$ of this calculation shows the maximum
         effect of pairing.
         Crust neutrons are always assumed to be paired in $^1S_0$ state with $T_c$
         taken from Ainsworth, Wambach \& Pines (1989).
         Core neutron pairing has no effect at the early stage (age below $\sim$ 20 yrs)
         but (i) slows down the cooling during the neutrino-cooling era 
         (age up to $\sim 10^5 - 10^6$ yrs)
         by strongly suppressing the neutrino emission, and (ii) fastens the cooling later 
         during the photon-cooling era by suppressing the star's specific heat by about 75\%
         (see, e.~g., Page 1994).
         The two sets of four curves show the effect of the crustal 
         magnetic field on the cooling
         by its influence on the outer boundary condition.
         The four $T_b - T_e$ relationships are taken from figure~\protect\ref{fig:tbte}.
         During the neutrino-cooling era the surface temperature
         follows the evolution of the core temperature:
         a $T_b - T_e$ relationship which gives a higher $T_e$, for a given
         $T_b$, results naturally in a higher stellar effective temperature.
         During the following photon-cooling era the effect is inverted since the dominating 
         energy loss is from surface thermal emission with a luminosity 
         $L_{\gamma} = 4 \pi R^2 \sigma T_e^4 \propto T_b^{2.2}$
         (see equation~\protect\ref{equ:tbte})
         and a higher $T_e$ for a given $T_b$ results in a larger $L_{\gamma}$.
         Calculations were performed with the cooling code used by Page (1994)
\label{fig:cool}}
\end{figure}

\end {document}
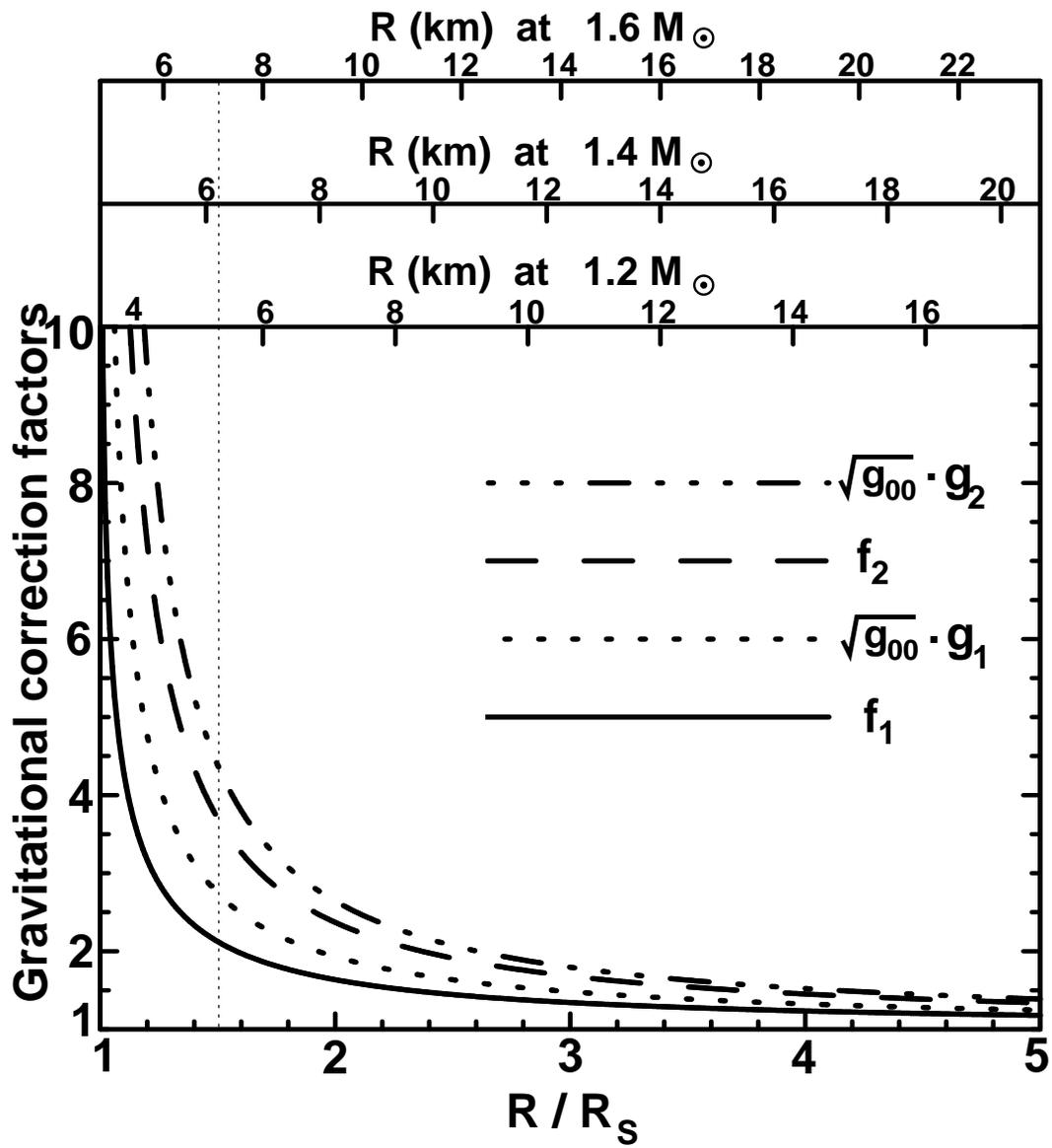
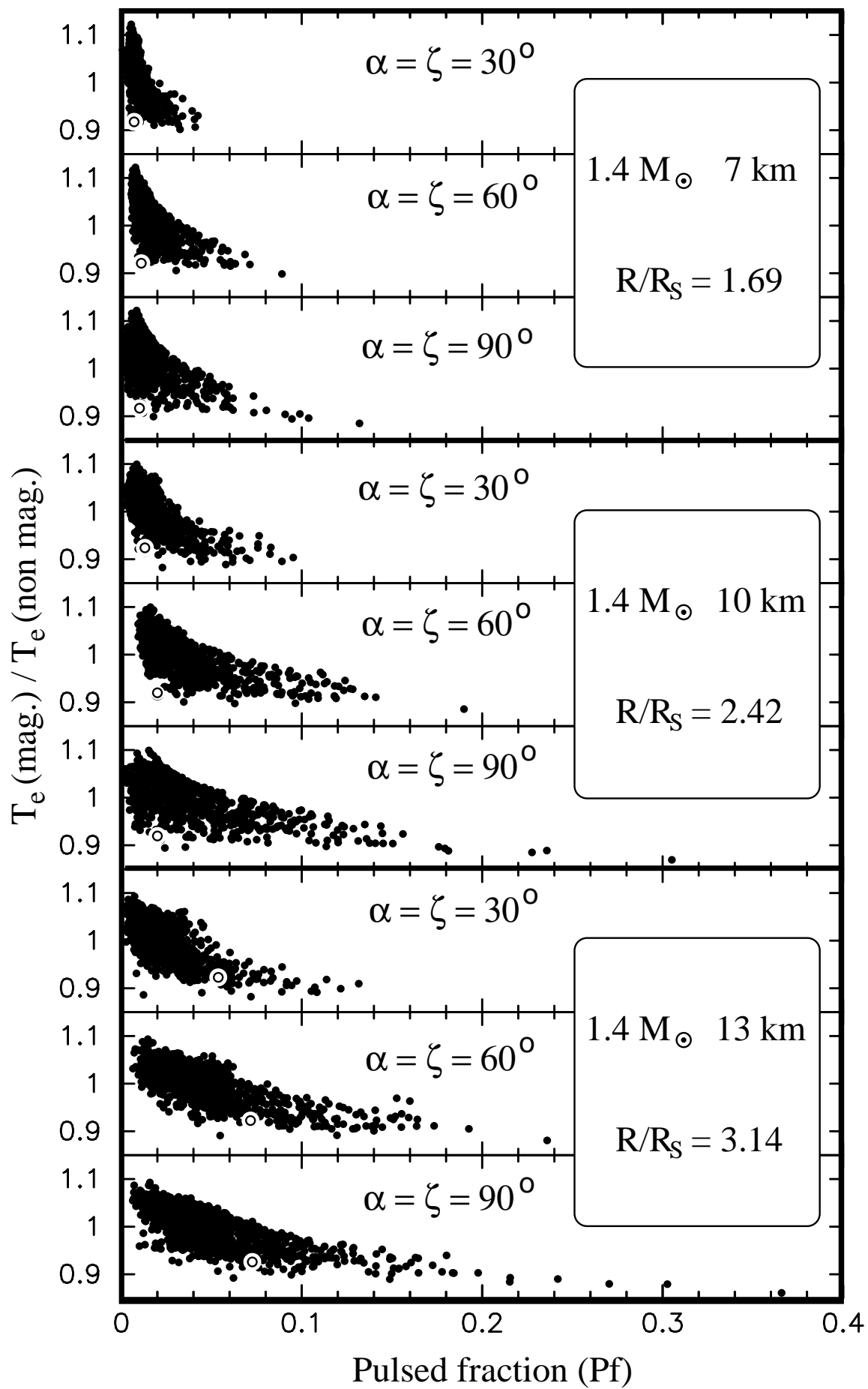
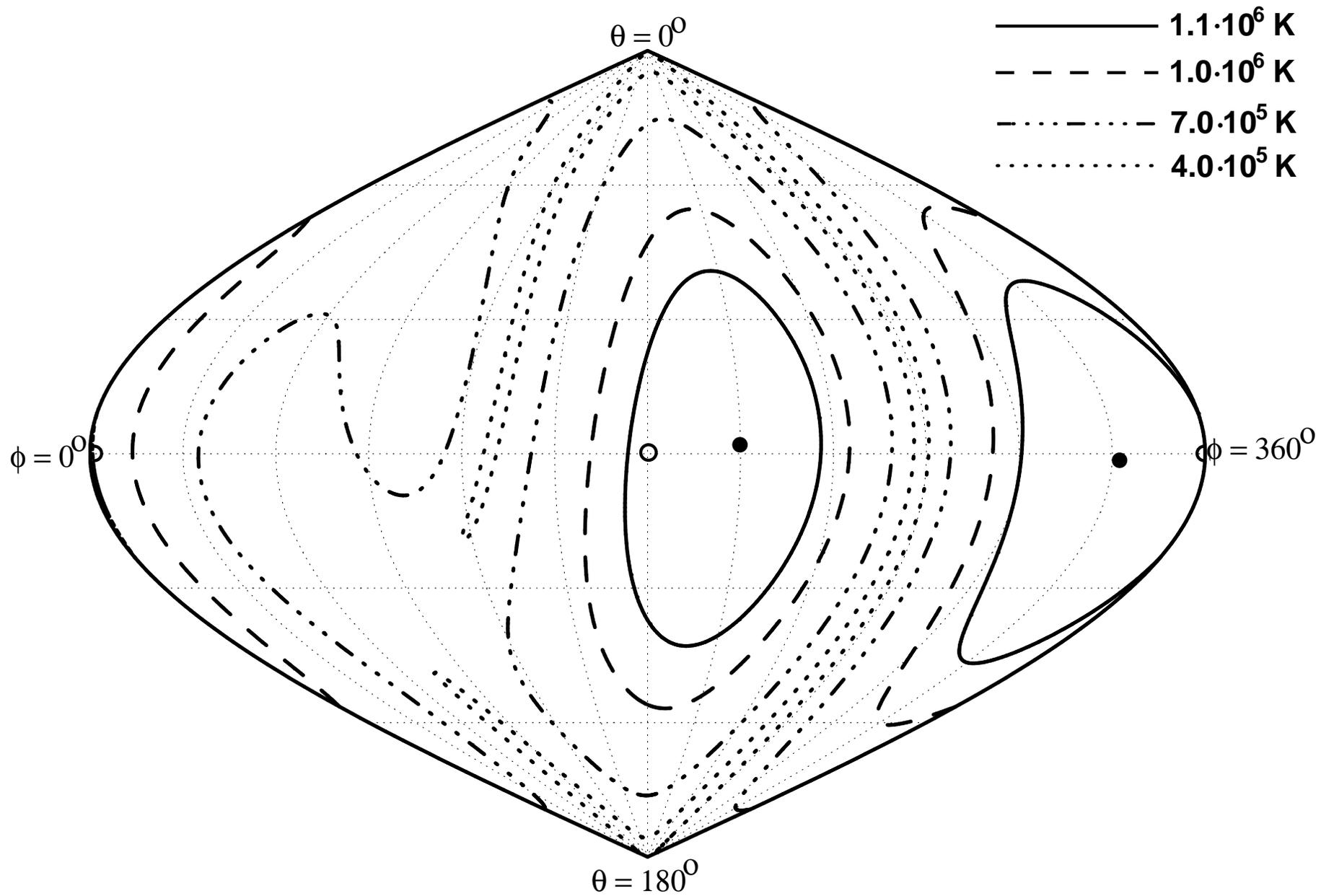
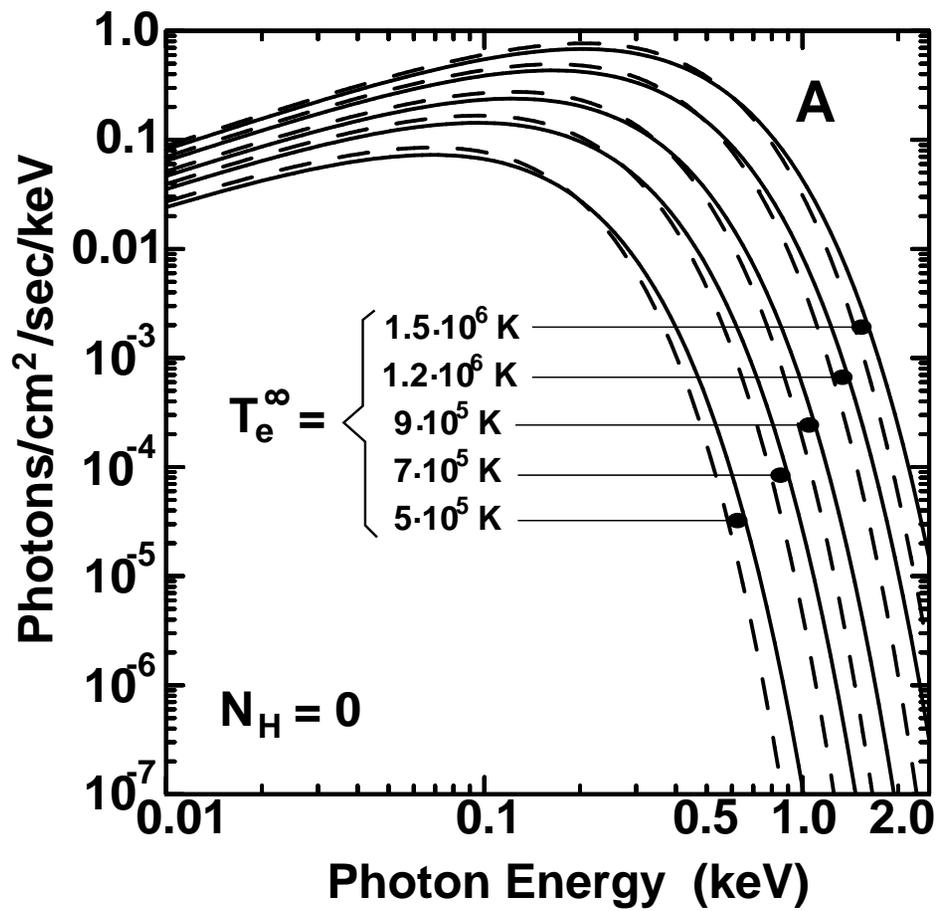
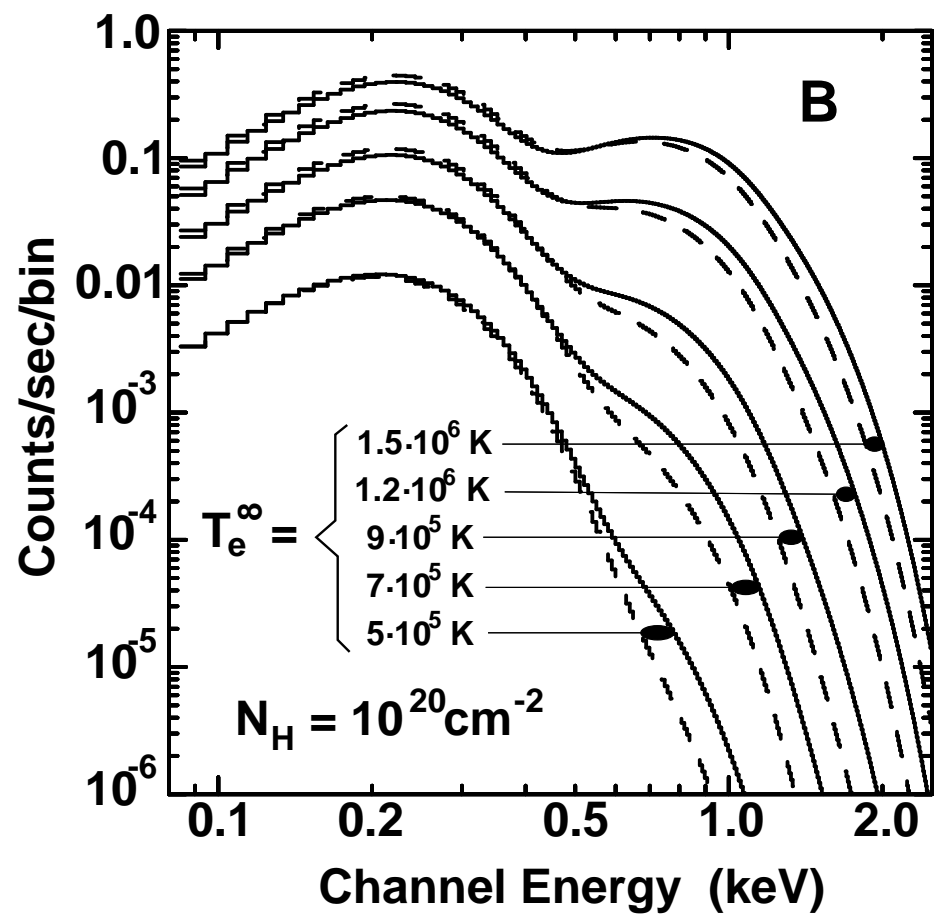
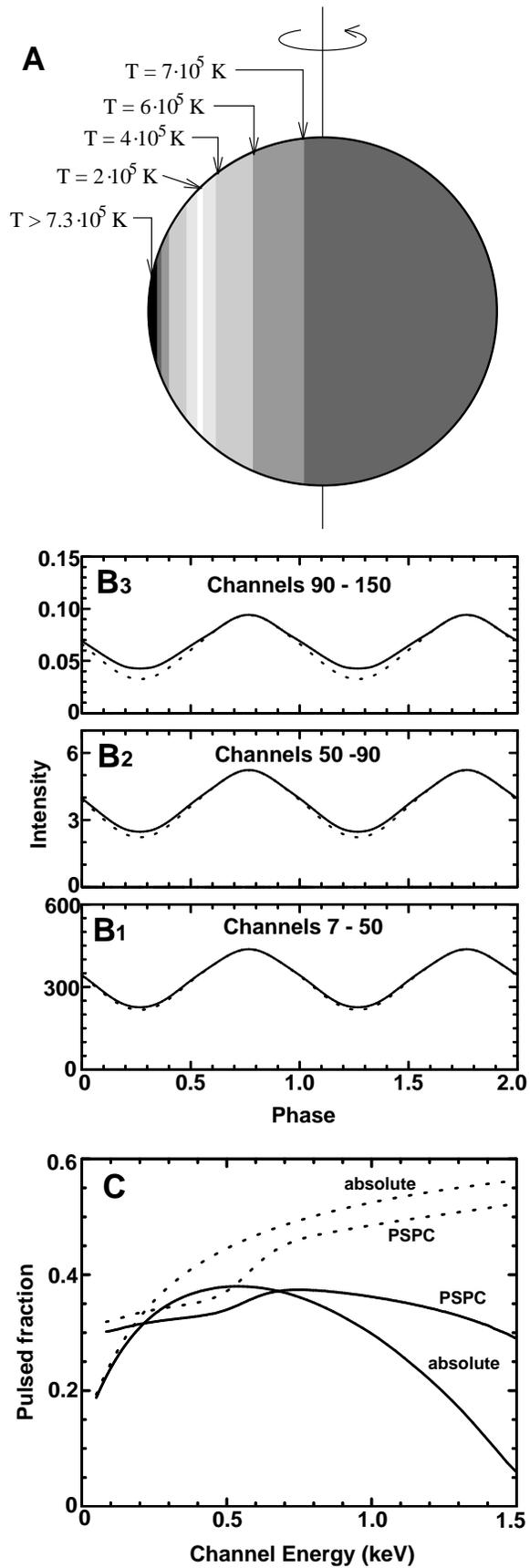
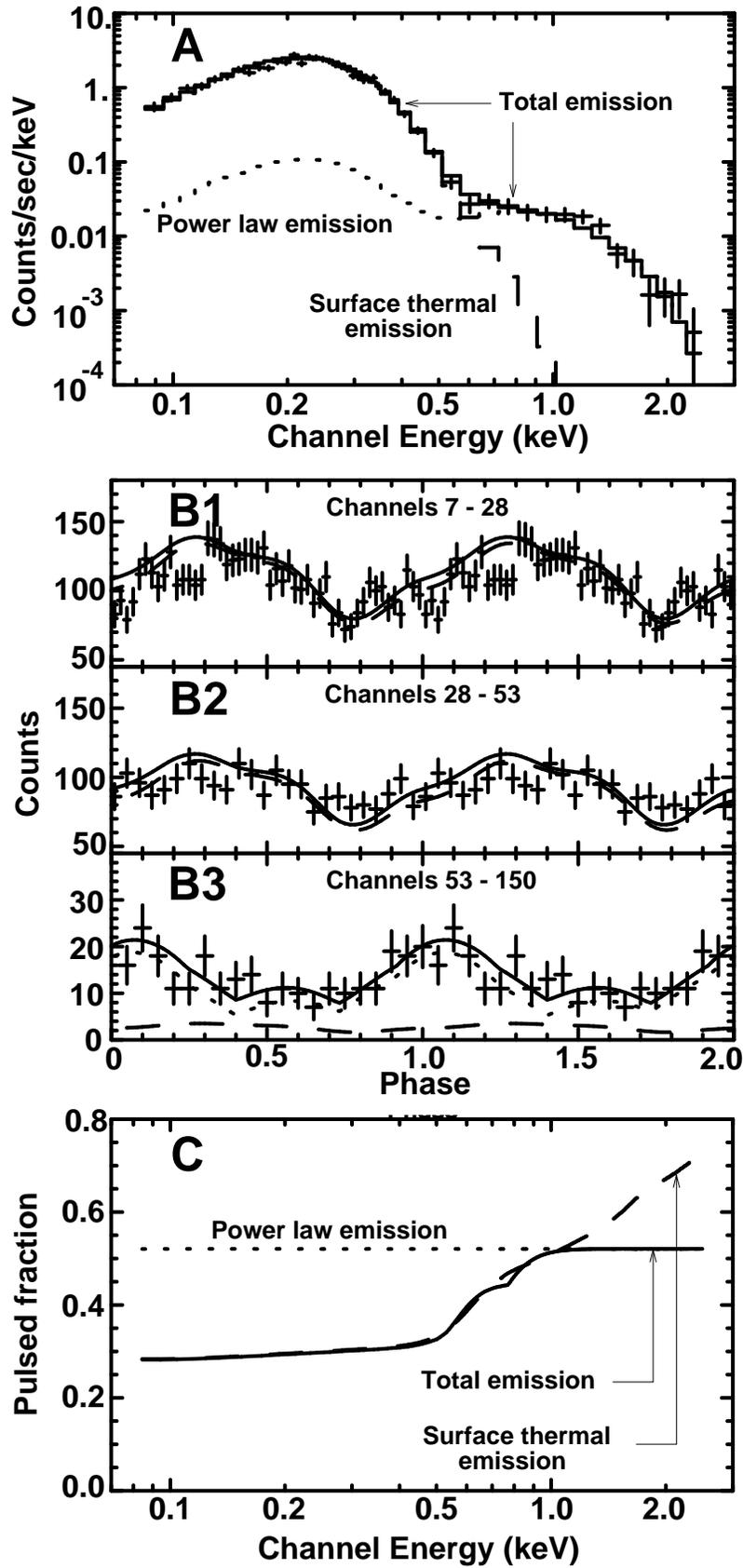
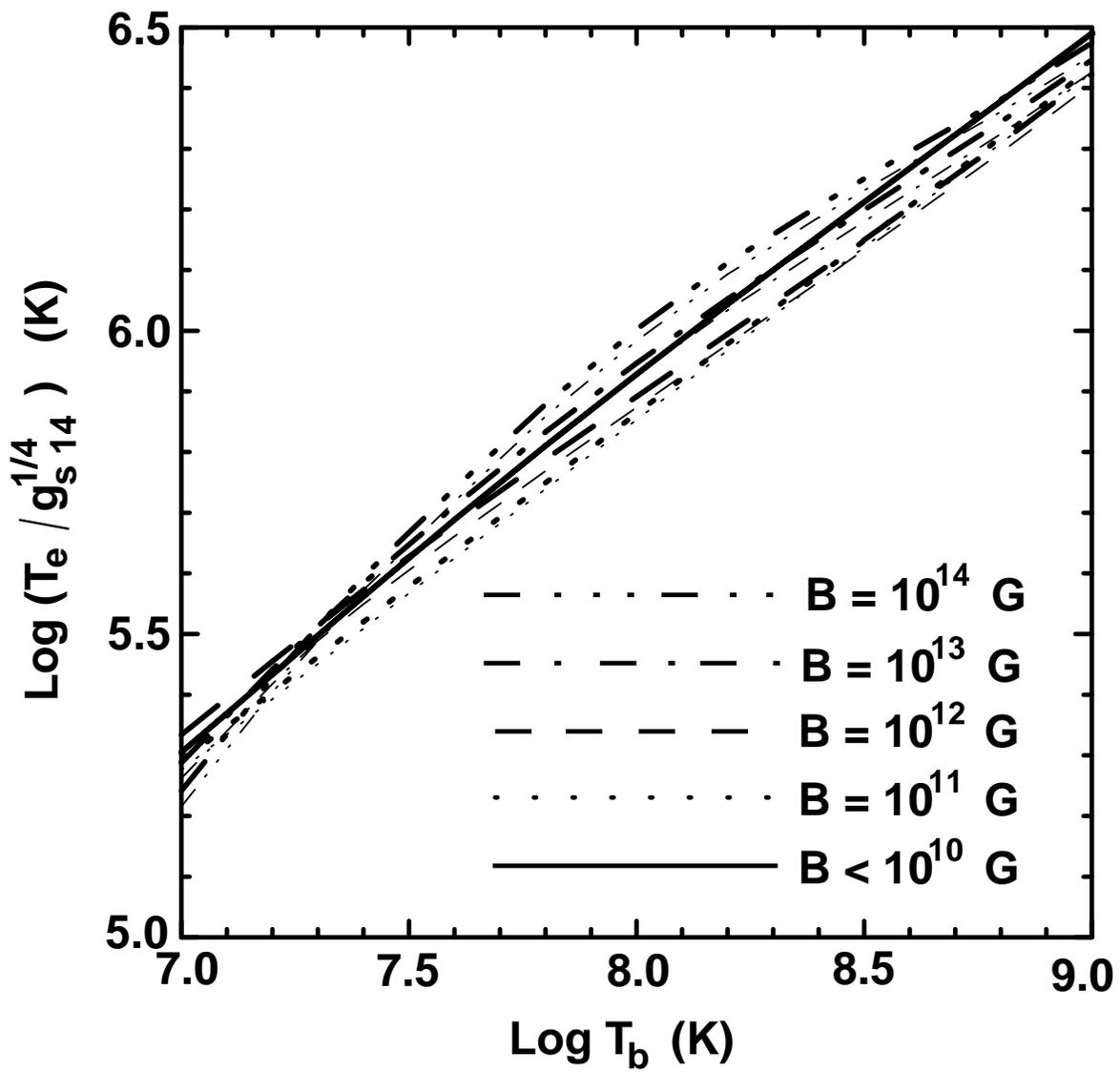
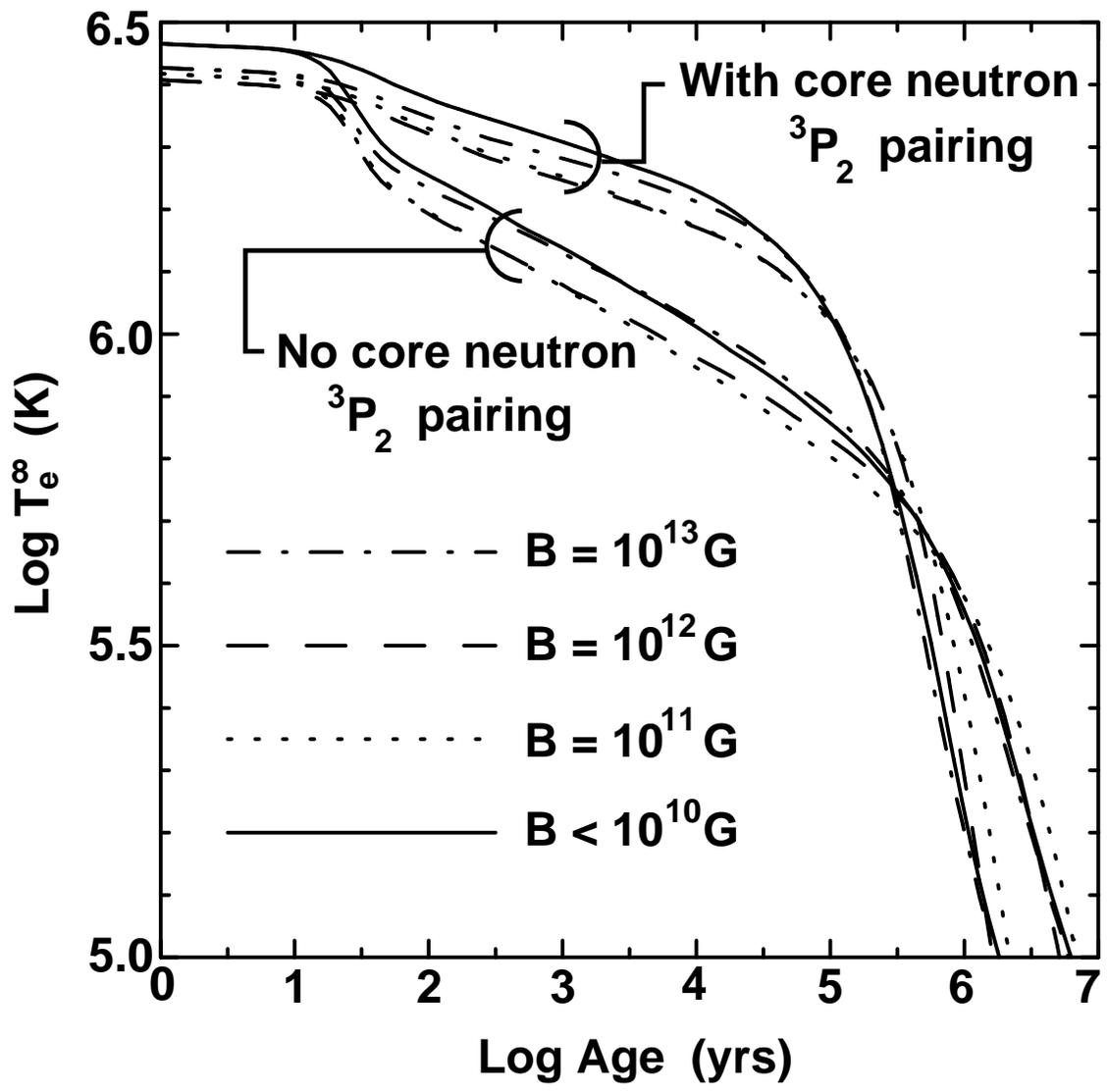
